# Detection of faint companions through stochastic speckle discrimination

Szymon Gladysz

*Department of Experimental Physics, National University of Ireland, Galway, University Road, Galway, Ireland*

szymon.gladysz@nuigalway.ie

and

Julian C. Christou

*Division of Astronomical Sciences, National Science Foundation, 4201 Wilson Boulevard, Arlington, Virginia 22230, USA* [1]

jchristo@nsf.gov

ABSTRACT

We propose a new post-processing technique for the detection of faint companions from a sequence of adaptive optics corrected short exposures. The algorithm exploits the difference in shape between the on-axis and off-axis irradiance distributions and it does not require the signal to be above the noise level. We show that the method is particularly useful in dealing with static speckles. Its application to real and simulated data gives excellent results in the low-signal regime where it outperforms the standard approach of computing signal-to-noise ratio on one long exposure. We also show that accurate noise estimation in adaptive optics images of close companions is rendered impossible due to the presence of static speckles. This new method provides means of reliable estimation of the confidence intervals for the detection hypothesis.

Subject headings: instrumentation: adaptive optics, methods: statistical

1. INTRODUCTION

The science task driving the development of the next generation of adaptive optics (AO) systems (the so-called extreme adaptive optics - ExAO) is the high-angular resolution imaging of very faint objects located close to their parent stars, e.g. exoplanets and brown dwarfs (Moutou et al. 2005). In order to obtain a $5\sigma$ detection of an Earth-like planet located at 10pc the required contrast is an overwhelming $10^{-10}$ at 0.1 arcseconds from the star (Cavarroc et al. 2006), and the integration time is about 100 hours using a 100-m telescope assuming the non-random errors are kept at a level of nanometres. This is indeed a formidable

---

[1] on leave from Center for Adaptive Optics, University of California, Santa Cruz, USA

task. The main source of error in the ExAO images will not be the background but the residual speckle noise (Racine et al. 1999) introduced by the atmosphere as well as the optical system. The static speckles produced by the manufacturing errors and less-than-perfect AO calibration will not average to zero over time (Hinkley et al. 2007).

In our previous work (Gladysz et al. 2006; Christou et al. 2006) we investigated the distribution of the instantaneous Strehl ratio (SR) with application to "lucky imaging" with AO. We discovered that the distribution of on-axis intensity in the AO-corrected short exposures has negative skewness, i.e. excess of low values. Negative skewness implies less high-quality outliers which could be eventually selected to form the shift-and-add (SAA) image. This means that "lucky imaging" is optimally used in the pure speckle or in the low-correction regime where the SR distribution has positive skewness or very small negative skewness.

Quite serendipitously, the negative skewness of the on-axis intensity distribution allows discrimination between the scattered speckles and real sources. The off-axis irradiance distribution is modelled by the modified Rician distribution (Aime & Soummer 2004; Soummer et al. 2007) which has positive skewness. We propose to use this asymmetry to discriminate between speckles – off-axis scattered light – and on-axis intensity coming from a faint companion. The new method would be optimally used on coronagraphic short exposures in the high-correction regime where static speckles dominate the noise. We emphasize that the technique is different from the "dark speckle" approach (Labeyrie 1995) because in principle it does not require the signal to be above the noise level.

In § 2 we describe the classical approach to detection in astronomy whereby the signal from a faint object is integrated until it exceeds the noise standard deviation by some factor. We also discuss the shortcomings of this method in the presence of static speckles. In § 3 we propose the new algorithm. We show its application to a binary star system imaged with the Lick Observatory AO system. The application of the method to simulated images of binary systems is discussed in § 4. We summarize its relevance to the field of high-contrast imaging in § 5.

2. CLASSICAL APPROACH TO DETECTION

In astronomy, peak-signal-to-noise ratio (PSNR) is almost exclusively used as the detection metric. There are two reasons for its popularity. Firstly, it is simple to calculate. Secondly, it gives reliable estimates for the probability of false alarm (PFA) and the missed-detection probability (MDP), given the noise variance and the expected signal level. PFA and MDP can be calculated because noise can be safely assumed to obey Gaussian statistics in the long-exposure regime, whatever the source of this noise is (background or speckles). Below is a brief and very simplistic description of the assumptions made – usually implicitly – in this detection scheme.

In the first case, a single star imaged against a sky background is treated. The long exposure of this star can be regarded as a sum of many independent short exposures, where both the background noise and the signal have Poisson statistics (ignoring scintillation and the addition of readout noise). By the central limit theorem (CLT), the resulting long exposure will have Gaussian statistics for both the signal and the noise. Therefore the PFA can be calculated analytically using the cumulative density function CDF of the noise process $I_n$:

$$\text{PFA} = 1 - \text{CDF}(I_t) = \frac{1}{2}\left(1 - \text{erf}\left(\frac{I_t - \bar{I}_n}{\sigma\sqrt{2}}\right)\right) \qquad (1)$$

where $I_t$ is the detection threshold ($I_t = \bar{I}_n + k\sigma$, $k = 1,2,3\ldots$), $\bar{I}_n$ is the mean noise intensity (usually zero), $\sigma$ is the noise standard deviation, and erf($x$) is the error function given by:

$$\text{erf}(x) = \frac{2}{\sqrt{\pi}} \int_0^x e^{-t^2} dt \qquad (2)$$

Using the above formula, one can calculate the PFA corresponding to the $5\sigma$ threshold to be $3 \cdot 10^{-7}$. The noise standard deviation is usually evaluated using an empty region in an image.

The MDP can be estimated in a similar fashion – given the expected flux from a star, the shape of the PSF, the pixel scale of the detector, etc. The variance of the signal is equated to its expected value as in the Poisson process. Then, one has a Gaussian process with an estimated expected value and variance. The MDP is the CDF of this new Gaussian process evaluated at the detection threshold $I_t$.

The second case is the situation where a star and its companion are imaged with AO. An astronomer wants to detect a faint companion "buried" in the light scattered from the bright star. Here, the main source of noise is the speckle contribution. With AO, the intensity of each speckle is distributed according to the modified Rician distribution (ignoring the Poisson and readout noise contributions). The star's peak signal has a negatively-skewed probability density function (PDF). Yet again, these characteristic distributions are smeared out according to the CLT and become Gaussian when long exposures are concerned. Equation (1) is also valid in this case.

The problem with the approach outlined above is not the validity of equation (1) but the process of noise estimation. The variance of speckle noise depends on its average value (Aime & Soummer 2004). If the PSF was isotropic there would not be a problem, but static speckles introduce anisotropy. These regions differ in mean brightness (and variance) so they represent random variables with different distributions. The assumption of ergodicity which is implicitly invoked when replacing ensemble averages by spatial averages cannot be used here. Now, because spatial moments of speckle intensity are no longer relevant then no constraints can be imposed on the brightness of static speckles. If one cannot say anything about the level of noise at the location of a suspected companion then, basically, PFA estimation is impossible.

The atmospherically-induced aberrations average to zero over time. Therefore, the detectability defined in terms of PSNR only depends on the static aberrations. These persistent speckles can be subtracted by a calibration procedure relying on the spectral properties of the faint object (Marois et al. 2005). In this approach however, there appears an additional problem of the non-common-path errors between the two spectral channels, which would introduce new static speckles. One can also image a point source immediately after the science target observation, but the variability of "seeing", other atmospheric parameters and the AO system's performance on a different target would affect the mean and variance of irradiance at the location of interest in the image. The classification of one feature as an object from among many static speckles can also be guided by some a priori information about the object's position (Janson et al. 2006).

We propose a new method, which we call stochastic speckle discrimination (SSD), allowing for the classification of object-like features in the AO-corrected images. The procedure has a major advantage in that it gives reliable estimates of the PFA without the need for the calibration techniques outlined in the previous paragraph.

3. STOCHASTIC SPECKLE DISCRIMINATION

3.1. The Rationale for Statistics-based Classification

The rationale for the new approach is that the distribution of the intensity of the PSF core (the faint companion) is different from that in the case of an off-axis speckle. This difference will be miniscule for faint sources, as the addition of intensities (from the companion, the speckles, the sky, etc.) implies the convolution of the individual PDFs. Therefore, the negative skewness of the on-axis intensity distribution will shift towards zero, and then will become positive, when the relative magnitude of the speckle and the sky contributions increases. One should not expect a negatively-skewed irradiance histogram at the location of the faint companion, because of the other random processes taking place at the same pixel.

We explained the negative skewness of the on-axis intensity distribution by modelling the effect of variable seeing (Gladysz et al. 2006) or changes in phase variance (Christou et al. 2006; Soummer & Ferrari 2007). In the latter approach, the normalized on-axis intensity, i.e. the Strehl ratio, was found to adhere to the following PDF:

$$p_{SR}(sr) = \frac{p_{\sigma_\phi^2}(-\ln sr)}{sr} \tag{3}$$

where $SR$ denotes the random variable with possible values $sr$, and $p_{\sigma_\phi^2}(\ )$ is the distribution of the phase variance described by the gamma model:

$$p(\sigma_\phi^2; k, \theta, \mu) = \frac{\left(\frac{\sigma_\phi^2 - \mu}{\theta}\right)^{k-1} \exp\left(-\frac{\sigma_\phi^2 - \mu}{\theta}\right)}{\Gamma(k)\theta} \quad \text{for } \sigma_\phi^2 \geq \mu \tag{4}$$

where $k > 0$ is the shape parameter, $\theta > 0$ is the scale parameter, and $\mu$ is the location parameter. The gamma function, $\Gamma(x)$, is given by:

$$\Gamma(x) = \int_0^\infty t^{x-1} e^{-t}\, dt \tag{5}$$

Equation (3) was obtained with the help of the Maréchal approximation which is only valid in the moderate to high SR regime:

$$SR = e^{-\sigma_\phi^2} \tag{6}$$

We followed the standard method of obtaining the PDF of a random variable, which has a functional relationship with another random variable with a known PDF (Goodman 2000). In Figure 1 the distribution of SR is plotted for three levels of turbulence strength, as quantified by the Fried's parameter, $r_0$.

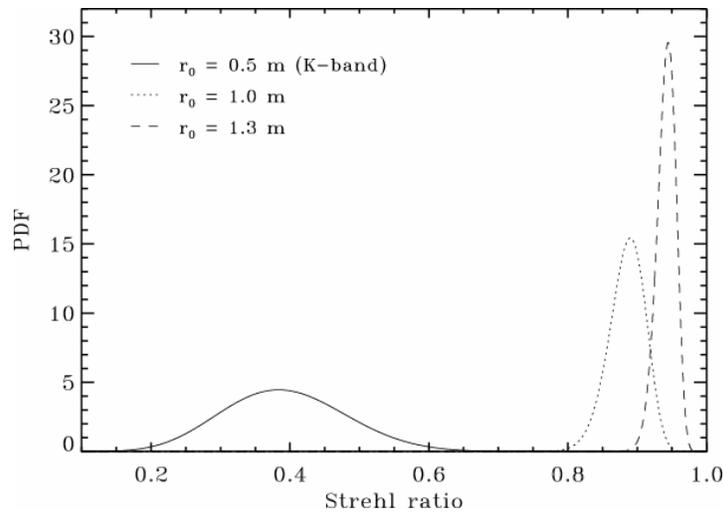

Fig. 1. Distributions of the Strehl ratio for three levels of turbulence strength.

On-axis intensity, which is linearly proportional to SR, is modelled by the scaled version of the SR PDF so it has the same shape. The off-axis intensity has statistics which follow the modified Rician PDF:

$$p(I) = \frac{1}{I_s} \exp\left(-\frac{I+I_c}{I_s}\right) I_0\left(\frac{2\sqrt{I}\sqrt{I_c}}{I_s}\right) \quad (7)$$

where $I_c$ corresponds to the intensity produced by the deterministic (constant) part of the wavefront, and $I_s$ corresponds to the halo produced by random intensity variations. $I_0$ is the zero-order modified Bessel function of the first kind. The parameters $I_c$ and $I_s$ are related to the expected value $E(I)$ and variance $\sigma_I^2$ of intensity through the following equations (Soummer et al. 2007):

$$E(I) = I_c + I_s$$
$$\sigma_I^2 = I_s^2 + 2I_c I_s \quad (8)$$

Figure 2 shows the modified Rician PDF for three levels of the static component $I_c$.

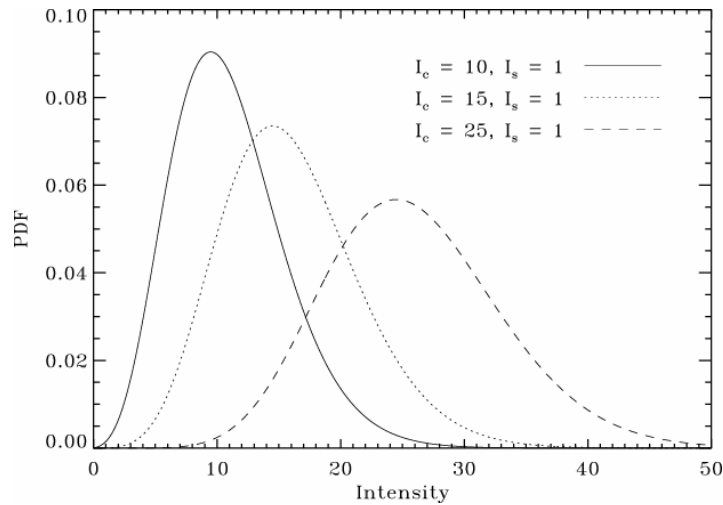

Fig. 2. PDF of speckle intensity for three different deterministic intensity levels $I_c$.

Comparing Figures 1 and 2 it is easy to see that in the moderate to high-SR regime the two PDFs have significantly different shapes. We use this morphological difference as a signal-noise discriminator.

It should be emphasized that the new method is based entirely on the dependence of the intensity statistics on the location within the PSF. Normally, the shape of the speckles and their brightness are employed in the discrimination process. Here, the location of the companion relative to the primary, whether it "sits" on a bright speckle, and their relative variances are equally important. In reality, the effectiveness of this new approach is expected to benefit from the same advantageous conditions as the standard PSNR analysis: further away from the core the speckles are fainter, their variances diminish, and the influence of the modified Rician PDF on the resulting composite statistics (companion + speckle) is smaller.

3.2. Speckle Discrimination Tested on Real Data

The binary star $\omega$ And (HD 8799, ADS 1152 AB) was observed with the Lick Observatory adaptive optics imaging system (Bauman et al. 1999) in order to test the method. The integration time was set to 22ms and we collected 10 000 exposures with a frame rate of 20Hz. The observations were carried out in the *K*-band, where the diffraction limit of the 3-m telescope is Nyquist-sampled by the detector pixels. For the details of the observations and data reduction see Gladysz et al. (2006). Figure 3 (left) shows the sub-pixel shift-and-add image of the binary star clearly showing the companion as well as the diffraction pattern of the Shane 3-m telescope including the effect of the spiders. Also note the residual speckles. The diffraction-limit $\lambda/D = 151$ mas.

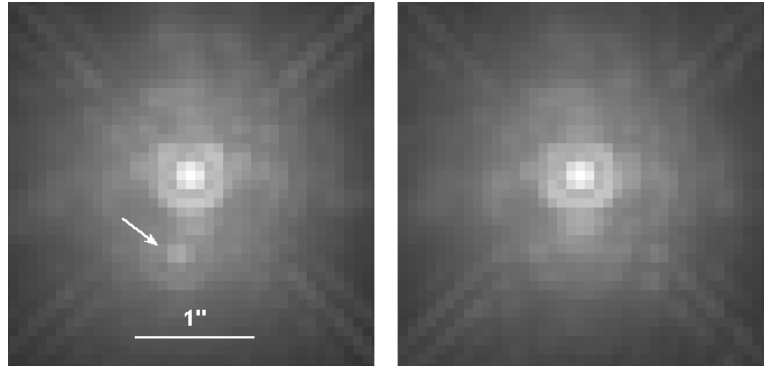

Fig. 3. *Left*: SAA image of the binary star $\omega$ And. The companion's location is denoted with an arrow. *Right*: SAA image of the PSF calibration star SAO 55171. Both images are displayed on a logarithmic scale. Note the similarity in the fixed speckle distributions.

The StarFinder PSF-fitting algorithm (Diolaiti et al. 2000) was used to extract relative photometry and astrometry of the system. The application of the code was problematic due to the lack of a calibration PSF. We used an SAA image of the star SAO 55171 (also shown in Figure 3), observed almost an hour before the science target. SAO 55171 is almost four magnitudes fainter in the visible, therefore the AO loop was running slower than during the binary observations; the SR (judged qualitatively) was also smaller. The integration time was set to 57ms (as opposed to 22ms for $\omega$ And) – this resulted in smaller readout noise. All of these effects translated to a PSF which had a slightly different morphology than the PSF from the later observations. StarFinder was designed to avail of many PSF estimates present in crowded fields. Here, we have a PSF estimate from a different time, different setup, and different atmospheric conditions. Therefore the results have to be treated as approximate.

Given the above concerns, it was surprising to find that the shape of the two PSFs (outside of the region contaminated by the light from the companion) was very similar. We started the algorithm with a rough guess for the position of the companion. The measured magnitude difference and separation were $\Delta m_K =$ 3.64 and $\theta = 0.67"$, respectively. We found that the algorithm had low tolerance to the accuracy of the initial guess: it would fail for estimated locations five pixels away from the true position. It also failed when SAA images of other single stars imaged that night were used to estimate the PSF.

In the simplest implementation of the SSD algorithm, the observer plots the intensity histograms at the location of the suspected companion, as well as at the locations of speckles of similar mean intensity (for the modified Rician distribution variance and skewness depend on the mean value). If the candidate is actually a speckle then all histograms will look similar. In Figure 4 four histograms are shown corresponding to intensities measured at the signal location (*a*) and three speckle positions (*b*, *c* and *d*). Light coming from the companion was focused between four pixels, and the mean value of these pixels was taken to represent the signal. The histogram did not change significantly when only the brightest pixel was used. Histogram (*a*) was therefore generated from the mean 4 pixel intensity over the complete data set of 10 000 frames.

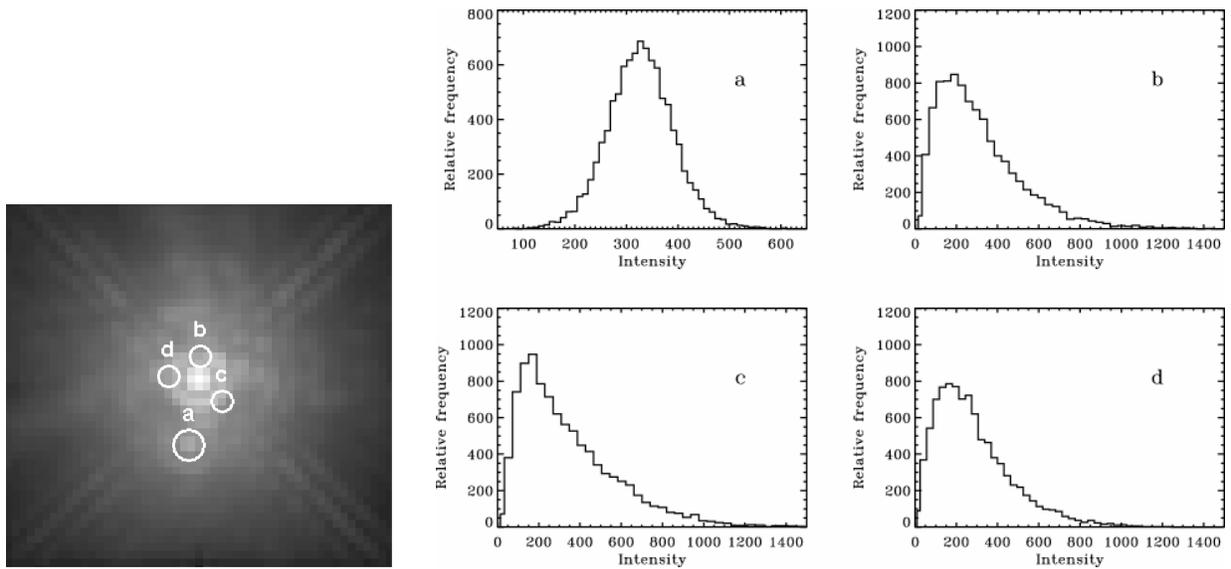

Fig. 4. Histograms of intensity at the locations corresponding to the companion (*a*), and the static speckles (*b*, *c* and *d*).

Clearly, the statistics of companion's intensity are different from the statistics of the speckles. The negative skewness is not visible anymore, but its influence on the distribution of the composite random process (intensity from the companion "sitting" on the speckle from the bright star) remains. It explains the symmetric histogram for position (*a*) – negative skewness has pushed the overall PDF towards a Gaussian-like shape. Convolution of two PDFs with equal variances but opposite skewness results in a symmetric PDF. For the moderate SR observed at Lick (0.4 – 0.6), the negative skewness of the signal is small compared to the positive skewness of noise. This implies that the variance of the signal must have been larger than the noise variance. This was to be expected – the signal is well above the noise level (average value in the surrounding pixels). Considering only Poisson statistics, if the mean of the signal is higher than the mean of the noise, the variance of the former must also be larger than the variance of the latter.

The above brief discussion opens an interesting avenue to develop an estimation technique. The method could be described as "PDF deconvolution". Two-dimensional image deconvolution could be replaced by a one-dimensional time-series deconvolution. The intensity distributions at the location of a star and a companion are scaled and shifted versions of the same SR PDF given by equation (3). This fact could be exploited in a one-dimensional, "blind", iterative deconvolution scheme with the modified Rician PDF as the convolution kernel. This approach is different from the "probability imaging" method (Carbillet et al. 1998) in that it exploits the morphological difference between the on-axis and off-axis single-pixel intensity distributions. The idea of "PDF deconvolution" will form the basis of our future work. In this paper we focus exclusively on the classification problem.

4. SPECKLE DISCRIMINATION APPLIED TO THE SIMULATED IMAGES OF THE BINARY SYSTEMS

4.1. Two-sample Goodness-of-fit Testing

In the previous section the difference between two intensity distributions was described qualitatively. For the systems with large magnitude differences the two histograms – one corresponding to the suspected source and the other corresponding to the "calibration" speckle statistics will differ only slightly. Moreover the binning intrinsic in any histogram computation will introduce extra ambiguity. There is a need to find an algorithm capable of quantifying the difference between two distributions. The algorithm should rely on raw (not binned) data.

There are a number of goodness-of-fit tests which utilize the concept of the empirical distribution function (EDF) – for an introduction see Wall & Jenkins 2003. The tests quantify the resemblance of the EDF – which is the proportion of the observed measurements that are less than or equal to a particular value – to the hypothesized cumulative density function (CDF). The computed test statistic is some measure of a distance between the EDF and the CDF. If the null hypothesis – that CDF is the underlying distribution – is correct, EDF should be close to CDF and this distance should be close to zero. The result of the test, the *p*-value, gives the probability that a value of the test statistic at least as large as the one observed, would have occurred if the null hypothesis were true. The greater the *p*-value the more confidence one can have in the null hypothesis.

The above discussion applies to a one-sample test, where the observations are hypothesized to obey some completely specified distribution. This approach is irrelevant here, as we cannot propose model Rician statistics for each point in the focal plane. Instead, we rely on two-sample tests wherein the second sample represents the calibration statistics. These tests are non-parametric, i.e. no assumptions about the population distribution (CDF) are needed. This alleviates the need for a bootstrap simulation to estimate the *p*-value, as the test statistic has a known – though usually complicated – sampling distribution. In our case, this second sample is taken from the data cube at the position of a static speckle with similar mean intensity to the suspected source. Obviously, this approach would fail if there were two companions of similar brightness in the system but this is highly improbable. In this procedure there is an additional risk of producing false alarms by mismatching two speckle distributions with the same mean intensity but different sets of parameters $I_c$ and $I_s$. We therefore implemented an estimation of these parameters through the method of moments (inversion of equation 8). Of course, at the location of a companion estimated parameters $I_c$ and $I_s$ did not correspond to the underlying speckle statistics, because the sample was a sum of two random processes. This sample was then compared via a two-sample goodness-of-fit test to the speckle time series with similar values obtained for $I_c$ and $I_s$. This approach resulted in the same number of false alarms as the first method where we only compared the means (see § 4.3). This was probably due to the error in parameter estimation through the method of moments. We therefore reverted to the original, simpler algorithm. We noted though, that $I_c$ and $I_s$ estimated for the companion's location were correspondingly lower and higher than the values obtained for speckles with similar mean value. This behaviour was very consistent in real data and simulations and we plan to exploit this effect in new "speckle discrimination" methods.

In this paper we use the Anderson-Darling two-sample test (Scholz & Stephens 1987), which measures the integrated squared difference between the two EDFs. As such, it is more powerful than the supremum-based tests, where the test statistic is the maximum distance between two EDFs (e.g. Kolmogorov-Smirnov test). The A-D test takes into account *all* the measured information. Moreover, the weighting function used in the integration makes this test more sensitive to the distribution's tails where we believe the departures from Rician statistics will be most evident.

The A-D test's *p*-value signifies the probability that the two samples came from the same population. For small values of this criterion, when the hypothesis of equal distributions must be rejected, it could be regarded as the probability of false alarm. If one is to make a detection decision, and this decision is based on the very low *p*-value, then this *p*-value could be treated as the PFA measure. Through simulations, the PFA resulting from the application of the A-D test to the short-exposure images was compared to the PFA from the standard, long-exposure approach.

4.2. Description of the Simulations

One approach to testing the performance of SSD for magnitude differences larger than ω And would be to use real, point-source data and scale and shift it to produce images of artificial binary systems. We have noticed though, that on-axis intensity statistics have more negative skewness and smaller variance for higher AO correction (Christou et al. 2006). The former effect – which would benefit the new method – cannot be re-created using real data from our Lick observations. We have therefore decided to employ Monte-Carlo simulations to generate data with the desired statistical properties.

Parameters of SPHERE, the planet finder instrument planned for the Very Large Telescope (Beuzit et al. 2006) were used in the PAOLA AO simulation package (Jolissaint et al. 2006). The coronagraphic module of the planned system was not included in our model. In the simulations the deformable mirror of the AO system had 40 actuators across its diameter. The imaging wavelength was 2.2$\mu$m, and at this wavelength the binary systems we modelled consisted of the primary star of magnitude 3, and the companion whose magnitude was varied between 13 and 16. It should be noted that the predicted dynamic range of this system is larger than the values reported here due to the light suppression by the coronagraph which we did not model. The sky brightness was set to 13 mag arcsec$^{-2}$. The atmosphere was assumed to consist of three dominant layers at the altitudes of 1, 5, and 10 km with wind speeds in these layers of 10, 15, and 25m/s. The seeing angle (corresponding to 0.55$\mu$m) was set to 0.5". The optical transmission of the entire system in the K-band was assumed to be 15%, filter bandpass was 0.32$\mu$m, quantum efficiency of the detector was 70%, gain was set to 10, and the integration time was 22ms. These parameters correspond to our imaging campaign at the Lick Observatory (Gladysz et al. 2006), only here the simulated detector had 1024 × 1024 pixels, and the focal-plane sampling was $\lambda/4D$, where $D$, the diameter of the VLT, is 8.2m.

PAOLA is an analytic code – it works by generating the AO-corrected phase structure function, which can be subsequently manipulated to obtain the long exposure. Since we wanted to obtain many short exposures we extracted the transitional output of the AO phase power spectrum corresponding to the SPHERE setup outlined above. This spectrum was then used as a filter to generate independent realizations of the AO-corrected phase. In order to include static speckles in the final images, we added to each phase a static component – the error map of the VLT's primary mirror[2] shown in Figure 5. This map was first scaled to have the rms phase variation of 5nm – this value is within the target range for future advanced AO systems (Cavarroc et al. 2006). The global tip and tilt were subtracted from the composite phase, the result was converted to electric field, multiplied by the binary mask representing the un-aberrated pupil, and the instantaneous PSF was generated by taking the squared modulus of the field's Fourier transform. The global linear components of the phase were subtracted because we wanted each pixel in the data cube to correspond to only one location in the PSF. The equivalent registering operation can be performed on the frames, and it would be made easy by the presence of a very bright source.

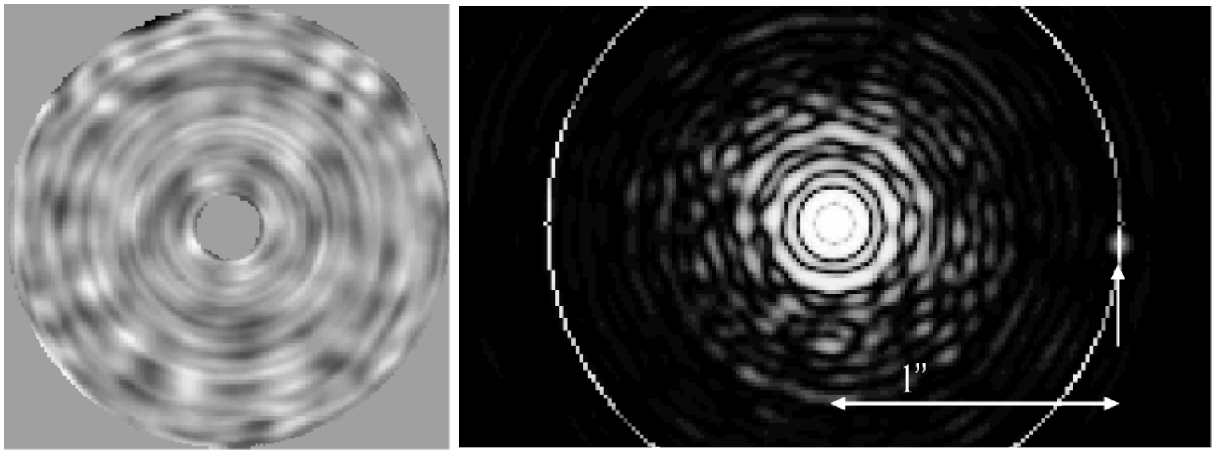

Fig. 5. Grey-scale image of the VLT mirror error map (*left*), and an SAA image of the binary system with magnitude difference of 8 in K-band (*right*); the arrow denotes the location of the companion. The white circle denotes the location of pixels which were used in the noise estimation process (in the program the pixels containing the light from the companion were excluded from this annulus).

The instantaneous PSF was copied and scaled to produce the desired magnitude difference. It was then shifted approximately 1" away from the centre of the frame. Two locations for the companion were investigated: one corresponding to a zero of the bright star's intensity pattern, and the other corresponding to a static speckle. As can be seen in Figure 5 the circular symmetry of the mirror's error map has translated into circular symmetry of the accumulated SAA image. It must be noted that the simulations did not generate on-axis time series with skewness as negative as encountered in our observations. This means that the results presented in § 4.3 are conservative compared to what might be possible in reality.

---
[2] http://www.eso.org/projects/vlt/unit-tel/m1unit.html

Noise-less short exposures were stored (eventually they were used to construct the SAA image). Noise was added to their copies in the following manner: first a sky frame was generated and added to the image, the result was contaminated by Poisson noise on a pixel-by-pixel basis. Finally, readout noise was added. Three readout noise levels were used throughout the simulations: 1, 5, and 10$e^-$.

The intensity time series corresponding to the pixel containing the peak of the companion's PSF constituted the signal. Then locations within the bright star's PSF with similar (± 5%) mean intensity were found. They were then sorted in the order of departure from the signal's mean value, and the neighbouring locations were discarded from future analysis. These procedures were done automatically without any human supervision. First three locations on the "calibration speckle" list were used as fake objects to determine how many false alarms SSD would produce. The three locations listed next had their time series combined and this way the calibration statistics were obtained. The companion's statistics, as well as fake source's were then compared to the calibration speckle statistics via the A-D test. The test allows for non-equal sample sizes, so we decided to use a large calibration sample in order to smooth out any departures from speckle statistics. It was observed that combining more than three time series to produce the calibration sample did not have a noticeable impact on the results. What was important was the total number of simulated frames – the length of each time series. For this particular setup SSD turned out to be unreliable for less than 3000 measurements. Most of the simulations were carried out for 5000-frame data cubes.

All time series were normalized by subtracting their mean value and dividing by the standard deviation. In the initial debugging runs the histogram of peak intensity of the bright star was also stored. This histogram would always display significant negative skewness, while the histogram of the companion's peak intensity had positive skewness just like the speckle histograms due to the convolution of the PDFs. The EDFs were computed for the normalized time series of the companion and the fake sources; then the A-D test was carried out between them and the calibration EDF. Since offset between the two time series was not relevant to this study (only the shape of the distributions) any residual bias of the time series was iteratively subtracted as to minimize the distance between the two EDFs. This procedure affected the sampling distribution of the test statistic and we recommend it only be used when the distribution of the A-D distance is estimated from the entire data-cube. Nevertheless, it must be said that the iterative bias subtraction did not change the A-D distance significantly. This is because the initial normalization was very accurate with such large samples. The Kolmogorov-Smirnov test was also used in the simulations but it produced results much inferior to the A-D test, and these results are not cited here.

The typical EDFs for the pixel containing light from the companion and for a static speckle are shown in Figure 6 together with the calibration distribution. As can be seen in the figure, negative skewness of the on-axis intensity distribution translates into the EDF which lies above the speckle distribution in the regions approximately one-standard-deviation left and right of the mean. This observation was used as an additional constraint in the algorithm: if a particular pixel's distribution did not have this property then it was automatically assigned to the speckle class. On many occasions fake sources were very quickly properly classified using this constraint.

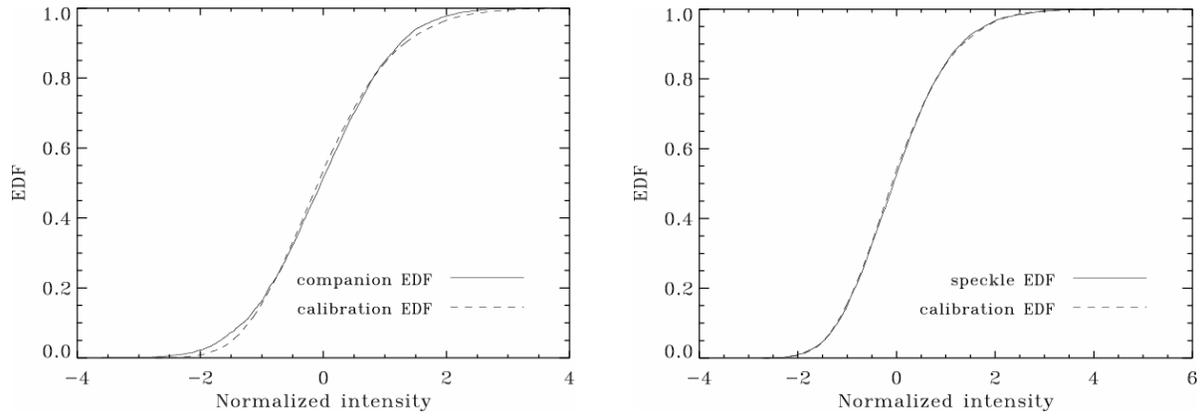

Fig. 6. Cumulative distribution of intensity at the location of the companion compared to its calibration EDF (*left*), and EDF for a static speckle of similar intensity as the source plotted with the same calibration EDF (*right*).

As was mentioned before, the SAA image was built as a sum of noise-less frames. It was subsequently contaminated with the sky noise corresponding to the duration of one simulation run (5000*0.022 = 110sec). Poisson noise, as well as a single realization of the readout noise were also added. This way the standard, long-exposure imaging could be compared to the short-exposure approach, where *each frame* is contaminated with the readout noise. The PSNR was computed on the SAA image in the following manner. The "signal" coming from the faint star was compared to the standard deviation of noisy speckle pixels located the same distance from the bright star as the companion. This is illustrated in Figure 5. It was assumed that the faint companion's PSF was located atop the speckle halo. The level of this halo decreases with distance away from the bright core. In the SAA image the measured peak intensity of the companion was reduced by the magnitude of the speckle halo estimated in the ring. The resulting signal was divided by the standard deviation of pixels in the ring:

$$\text{PSNR} = \frac{I_{peak} - \bar{I}_{ring}}{\text{stddev}(I_{ring})} \qquad (9)$$

In the above equation the denominator represents mainly the static noise contribution. Of course, the contribution from sky and instrument noise was also present on top of the speckles, but these effects were negligible in this situation. The PFA corresponding to PSNR from equation (9) was calculated with the help of equations (1) and (2) assuming Gaussian statistics. Following the discussion in § 2, we have to state that the measurement of noise outlined above is not rigorous, but we believe that noise cannot be estimated rigorously in the presence of static speckles.

Each simulation run for a particular choice of parameters (magnitude difference, location of the companion, and the readout noise level) was repeated ten times. Median values of PFA resulting from the short- and long-exposure methods were stored, together with the number of the SSD false alarms, i.e. situations when the A-D test's *p*-value was below 0.05 or 0.01 for pure speckles. It should be mentioned that the resources that the simulations called for were beyond the capabilities of a desktop PC – the code was executed in parallel on the 32-node Hobbes cluster in the NUI Galway IT Department[3]. We also used the 476-node Walton cluster, part of the Irish Centre for High-End Computing[4]. Each simulation run took approximately a week.

4.3. Results of the Simulations

---

[3] http://www.scg.nuigalway.ie

[4] http://www.ichec.ie

Figure 7 shows the PFA for the new method, contrasted with the PSNR PFA. The diagonal line denotes perfect equivalence of the two methods. The right halves of each plot correspond to large magnitude differences (weak signal – large PFA), the left halves correspond to relatively stronger signals. Points below the diagonal line indicate cases when SSD gave better results (lower PFA) than the standard approach.

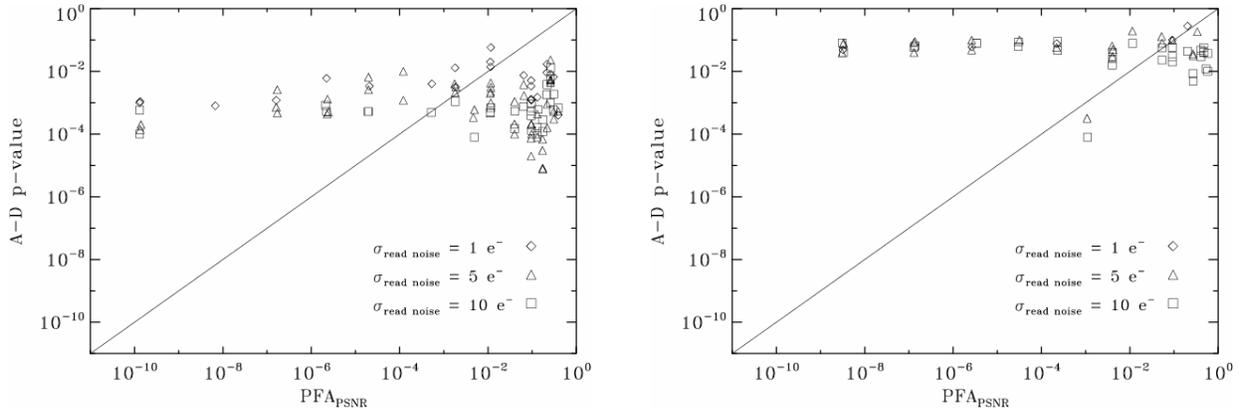

Fig. 7. The A-D test's *p*-value vs. the PFA associated with the PSNR for the companion located on the zero of the bright star's PSF (*left*), and when it was located on a static speckle (*right*).

SSD performed better when the companion was not located on a static speckle – the points plotted in the left panel in Figure 7 are generally lower than in the right panel. This is easy to explain with the knowledge of the Rician statistics. Static speckles have larger intensity variances than the low-level locations. This relatively large variance affects the composite statistics when the source is "sitting" on the speckle, diminishing the effect of the source's negative skewness (see the discussion in § 3.2). There is also an additional effect with a reverse action: low-level intensity has large positive skewness (asymptotically approaching exponential statistics), while static speckles have almost symmetric PDFs. This suggests that companion's statistics should be less affected by the static speckles than pure speckles. It seems though, judging from Figure 7, that speckle variance has a larger impact than speckle skewness.

Both panels in Figure 7 suggest that SSD performs better than PSNR in the low-signal regime. In some cases when PSNR was close to zero, detection was still made possible by the new method (*p*-value was less than 0.01).

An interesting observation is that there are no easy-to-identify trends in both plots in Figure 7. SSD PFA was expected to decrease as the signal got stronger (for smaller magnitude differences). The signal would then dominate the composite statistics and the EDF would be further apart from the pure-speckle calibration EDF. This effect cannot be seen in Figure 7 because we were mainly interested in weak signals and the effective range of PSNR was 0 to 10. The plot range was also limited by the accuracy of numerical approximation to the tail of the error function given by equation (2). This accuracy is only $10^{-9}$. To test the hypothesis that the *p*-value actually decreases for stronger signals we looked at a wider range of magnitude differences. In Figure 8 magnitude difference is plotted versus the A-D *p*-value. This time the trends are clearly visible. It must be said that the slopes of both curves are small, and therefore the applicability of the new method lies in the low-signal regime.

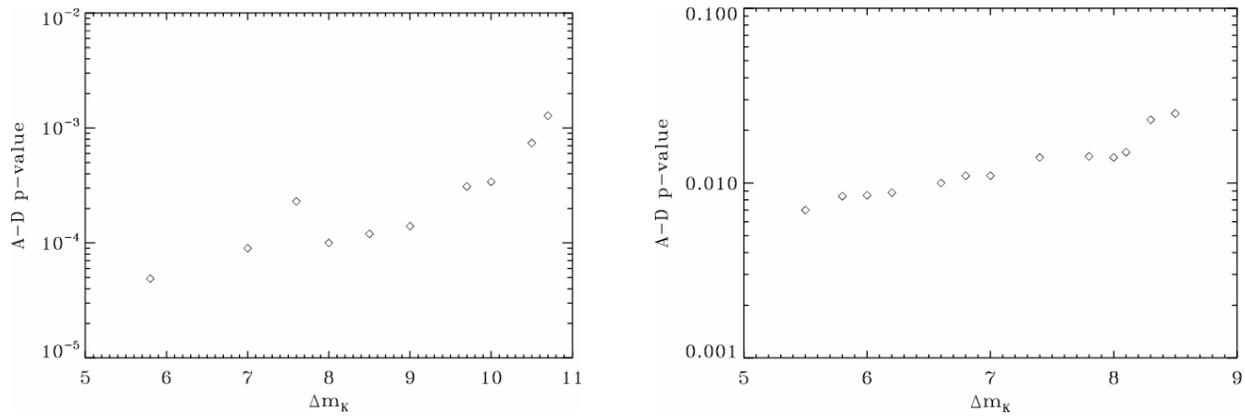

Fig. 8. The A-D test's *p*-value vs. magnitude difference for the companion located on the zero of the bright star's PSF (*left*), and when it was located on a static speckle (*right*).

SSD, as described above was also tested on $\omega$ And. A very small *p*-value was expected for the companion's time-series, and indeed the obtained value was equal to zero.

Speckle locations were also tested during the simulations to see if the A-D *p*-value truly gives the PFA. The percentage of false alarms for two confidence levels is shown in Figure 9. The graph contains the consolidated results from all simulation runs. The huge majority of runs did not produce any false alarms – there occurred only thirteen of them for very weak signals. As expected, there were fewer false alarms at the 0.01 confidence level than at the 0.05 level. Assuming one would set the confidence level to 0.01 in the actual observations, the SSD PFA should be around 10% as compared to almost 30% with the standard method (keeping in mind that this conclusion was drawn from just a few points).

But why were there more false alarms towards weaker signals in the first place? This trend would mean that the statistics at lower light levels are somehow more volatile. This is strange because the A-D distance has only one distribution, which should be independent of the moments of the investigated sample and its calibrator. The PFA captured by the A-D test's *p*-value can therefore be regarded more as a measure of confidence level with which one can announce detection. It seems that it does not correspond to the actual *rate* of false alarms at some light level. We expected this rate to be constant across the different signal strengths and equal to the confidence level of 0.05 or 0.01. One explanation is that at low-light levels the algorithm picks "empty" calibration pixels which are almost entirely Gaussian and compares them to low-level, exponentially-distributed speckle. This, of course, would result in a very low *p*-value. For real data we recommend that the distribution of the A-D distance be estimated from the entire data-cube. This system would include cases of aforementioned mismatch in the calculation of PFA.

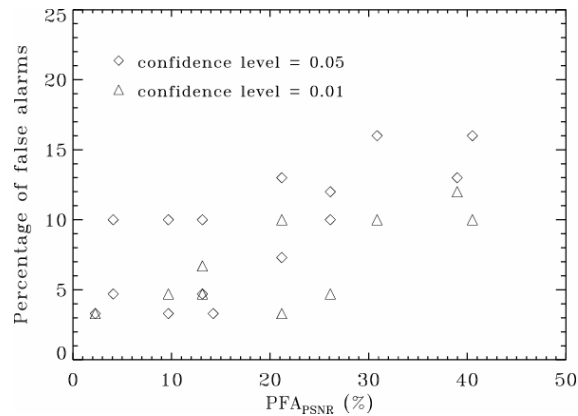

Fig. 9. Percentage of SSD false alarms vs. the PSNR PFA (measured on the faint companion). Only non-zero results are shown.

## 5. CONCLUSIONS

We have presented a new, statistics-based method of feature-classification in AO images. The new algorithm relies on a large number (> 3000) of short-exposure frames. We show that it performs favourably in the weak-signal regime against the standard method of summing all the frames, and computing the PSNR on the SAA image. Its advantage would be particularly clear when applied to cases where the SAA image contains several object-like features, like the one published by Janson et al. (2006).

In our opinion SNR estimation in AO images of close companions is rendered very inaccurate or even mathematically invalid by the presence of static speckles. The new method could provide reliable estimates of the PFA. In the future we plan to estimate the confidence intervals of the test statistic (the A-D distance) directly from the data cube. This process would do away with any inconsistencies in the distribution of the test statistic across the image. These inconsistencies would arise with real data, where intensity time series are not stationary and the asymptotic distribution of the test statistic – which was given analytically for stationary samples – would not be relevant.

One concern is that the method requires good AO correction to avail of the negative skewness effect. This means achieving SR of more than 0.4 (Christou et al. 2006), which is currently only possible in the infrared. Unfortunately there are no infrared detectors equivalent to the low-noise, fast-readout CCDs used in the visible band (Law et al. 2006). What is encouraging is the fact that SSD performs well even when the readout noise is quite high (10e$^-$) as can be seen in Figure 7. Some modern infrared detectors are capable of achieving fast frame rates with readout noise of a few photoelectrons (e.g. PHARO at the Palomar Observatory's 5-m telescope).

The observer wishing to implement described methods should be warned against using the shortest possible exposure times for fainter sources. Low-light level data will exhibit Poisson statistics in every pixel and speckle discrimination will not work. It is important to obtain sufficient number of photons in each pixel in order to avoid this effect. For bright stars, and relatively bright companions, exposures as short as milliseconds could be used (given detectors capable of such frame rates). For fainter objects, we believe exposure times as long as half a second in $K$-band will not make all the statistics Gaussian even though central limit theorem suggests otherwise. This is because the correlation time of on-axis intensity in the case of AO-corrected exposures is quite long, between 0.05 and 0.35sec in $K$-band, even though speckle "boiling" time is only tens of milliseconds (Gladysz et al. 2006). This means there would be almost no penalty (i.e. smearing of the statistics) if one chooses exposure times shorter than one second. With that in mind we present a very challenging case of trying to image an exoplanet with the new technique. Recently, Itoh et al. (2006) reported possible detection of an exoplanet 0.9" away from the star $\varepsilon$ Eri in an AO coronagraphic image taken with the 8.2-m SUBARU telescope. The detection is hard to confirm due to a large number of artefacts (static speckles) in the image. The feature has an estimated $H$-band magnitude of 17.3. At this brightness level, SUBARU collects – depending on the instrument and the pixel scale – between 10 and 100e$^-$ in the peak pixel during 0.3sec which was the exposure time used by the authors. This suggests that stochastic speckle discrimination could be used to detect even (giant) exoplanets.

In studying exoplanets and sub-stellar companions to bright stars it is almost always necessary to use short exposures to avoid saturation of the detector. We suggest the statistical information which is present in these frames be used in the detection process. It could complement the standard, PSNR-based method. Perhaps it would be desirable to confirm a suspected detection made via the PSNR, by taking a few thousand short exposures, and looking at the intensity histogram at the location of a candidate companion. It should also be mentioned that accumulating more signal does not necessarily benefit the PSNR in the presence of fixed speckles. This is predicted by other authors (Hinkley et al. 2007). In one of our simulation runs the PSNR changed from 9 (in a single frame) to 10 (when ten thousand frames were stacked together). This miniscule gain shows how difficult it is to fight fixed speckle noise by simple temporal integration.

SSD would not benefit from spectral differential imaging (Marois et al. 2005). The static PSF subtraction from every frame would not affect the intensity statistics apart from shifting the mean value. Speckle discrimination would however benefit from the inclusion of a coronagraph in the optical system. The coronagraph would reduce the mean intensity up to a certain point in the focal plane. This would reduce the probability of a companion lying on top of a static speckle of significant intensity. Figure 7 suggests that SSD would perform well in this case.

This research was supported by Science Foundation Ireland under Grants 02/PI.2/039C and 07/IN.1/I906, as well as the National Science Foundation Science and Technology Center for Adaptive Optics, which is managed by the University of California at Santa Cruz under cooperative agreement AST 98-76783. We would like to thank the staff of Lick Observatory, in particular Elinor Gates. In addition we would like to thank Nicholas Devaney for useful comments on the draft version of this paper, Rémi Soummer and Anand Sivaramakrishnan for discussions on speckle statistics, Granville Tunnicliffe-Wilson and Jerome Sheahan for the introduction to time series analysis, Michael Fitzgerald for information about the high-speed mode of the camera at the Lick Observatory, and Chris Dainty for support for this research.

The authors wish to acknowledge the SFI/HEA Irish Centre for High-End Computing (ICHEC) for the provision of computational facilities and support.